\documentclass[12pt,epsf]{article}
\usepackage{amssymb,amsmath}
\usepackage{graphicx}

\newcommand{\be}{\begin{equation}}
\newcommand{\ee}{\end{equation}}
\newcommand{\bea}{\begin{eqnarray}}
\newcommand{\eea}{\end{eqnarray}}
\newcommand{\beas}{\begin{eqnarray*}}
\newcommand{\eeas}{\end{eqnarray*}}
\newcommand{\ba}{\begin{array}}
\newcommand{\ea}{\end{array}}

\newcommand{\tr}{\mathrm{Tr}}

\def\identity{{\rlap{1} \hskip 1.6pt \hbox{1}}}
\newcommand{\nbox}{{\,\lower0.9pt\vbox{\hrule \hbox{\vrule height 0.2 cm \hskip 0.19 cm \vrule height 0.2 cm}\hrule}\,}}

\def\href#1#2{#2}

\begin{document}
\begin{titlepage}
\hfill

\vspace*{20mm}
\begin{center}
{\Large \bf Cold Nuclear Matter In Holographic QCD }

\vspace*{15mm}
\vspace*{1mm}

Moshe Rozali${}^1$, Hsien-Hang Shieh${}^1$, Mark Van Raamsdonk${}^1$, and Jackson Wu${}^2$

\vspace*{1cm}

{${}^1$Department of Physics and Astronomy,
University of British Columbia\\
6224 Agricultural Road,
Vancouver, B.C., V6T 1W9, Canada\\
\vskip 0.1 in
${}^2$ Theory group, TRIUMF,
4004 Wesbrook Mall,
Vancouver, BC, V6T 2A3, Canada}

\vspace*{1cm}
\end{center}

\begin{abstract}
We study the Sakai-Sugimoto model of holographic QCD at zero temperature and finite chemical potential. We find that as the baryon chemical potential is increased above a critical value, there is a phase transition to a nuclear matter phase characterized by a condensate of instantons on the probe D-branes in the string theory dual. As a result of electrostatic interactions between the instantons, this condensate expands towards the UV when the chemical potential is increased, giving a holographic version of the expansion of the Fermi surface. We argue based on properties of instantons that the nuclear matter phase is necessarily inhomogeneous to arbitrarily high density. This suggests an explanation of the ``chiral density wave'' instability of the quark Fermi surface in large $N_c$ QCD at asymptotically large chemical potential. We study properties of the nuclear matter phase as a function of chemical potential beyond the transition and argue in particular that the model can be used to make a semi-quantitative prediction of the binding energy per nucleon for nuclear matter in ordinary QCD.

\end{abstract}

\end{titlepage}

\vskip 1cm

\section{Introduction and Summary}
\vskip 0.1 in
\noindent
{\bf QCD at finite temperature and chemical potential}
\vskip 0.1 in
The phase diagram of QCD as a function of temperature and baryon chemical potential (or alternatively baryon density) displays a rich variety of phases and transitions (for reviews, see \cite{rw,schaefer,stephanov}). However, apart from the regimes of asymptotically large temperature or chemical potential, where some analytic calculations are possible, and of zero chemical potential, where reliable lattice simulations are possible, our knowledge of the phase diagram is based exclusively on extrapolations and semi-empirical toy models. For intermediate values of the chemical potential, numerical simulation is plagued by a notorious `sign problem' (see for example \cite{stephanov}), while analytic calculations are not possible due to strong coupling. Thus, while there has been significant progress recently in understanding the qualitative features of the phase diagram, reliable quantitative calculations that would definitively verify the proposed phase structure or determine the locations of various transitions or properties of the various phases seem a formidable challenge at present. A better understanding of the details of the phase diagram at intermediate chemical potential would have valuable applications, for example in understanding the physics of neutron-star interiors.
\vskip 0.1 in
\noindent
{\bf Holographic models of QCD}
\vskip 0.1 in
With the advent of the gauge theory / gravity duality \cite{maldacena}, we have a new tool for studying the properties of certain strongly coupled gauge theories. While the original and most studied examples involve highly supersymmetric conformal gauge theories without fundamental matter, much progress has been made in constructing examples without supersymmetry \cite{wittenthermal}, with confinement \cite{wittenthermal}, with fundamental matter \cite{kk} and with chiral symmetry breaking \cite{ks}. We now have examples of gauge theories with a known gravity dual that share most of the qualitative features of QCD, and the duality permits analytic calculations that would be otherwise impossible.

It is obviously interesting to study these QCD-like theories in regimes for which neither analytic or numerical studies are currently possible in real QCD. One such regime is the near-equilibrium behavior of the theory at finite temperature . This has received a great deal of attention recently (see \cite{sonstarinets} for a review) since calculations in holographic\footnote{Here 'holographic' is a now conventional term referring to the equivalence between a higher-dimensional gravitational theory and a lower-dimensional field theory.} models of QCD-like theories do a better job of explaining and predicting some properties of the quark-gluon plasmas produced in relativistic heavy-ion collisions than any other approach. In the present paper, our focus will be on another such regime as described above, the equilibrium properties at finite baryon chemical potential.

There is already a large literature on studies of gauge theories at finite chemical potential using gravity duals (see \cite{Kobayashi:2006sb} and references therein). Many of these consider a chemical potential for R-charge in theories with only adjoint matter. There have been some some studies of the behavior of theories with fundamental matter at finite baryon chemical potential, but the early examples of holographic theories with fundamental matter had both bosonic and fermionic fields carrying baryon charge. In these cases, the physics at finite chemical potential involves Bose condensation rather than the formation of a Fermi surface. In order to get behavior similar to real QCD, it is essential to study a theory with baryon charge carried exclusively by fermionic fields. Such a model was constructed a few years ago by Sakai and Sugimoto \cite{ss1}, and it is this model that we will focus on the present work.
\vskip 0.1 in
\noindent
{\bf The Sakai-Sugimoto model}
\vskip 0.1 in
The details of the Sakai-Sugimoto model are reviewed in section 2. Briefly, the model gives a holographic construction of a non-supersymmetric $SU(N_c)$ gauge theory with $N_f$ fundamental fermions. The gravity dual involves $N_f$ D8-branes in the near-horizon geometry of $N_c$ D4-branes wrapped on a spatial circle with anti-period boundary conditions for the fermions. In the geometry, the compact direction of the field theory together with the radial direction form a cigar-type geometry, in which the D8-branes are embedded as shown in figure 1. The other directions include an $S^4$ carrying $N_c$ units of D4-brane flux and the $3+1$ directions of the field theory. In addition to $N_f$ and $N_c$, the theory has a dimensionless parameter $\lambda$, the 't Hooft coupling at the field theory Kaluza-Klein scale.\footnote{The model has another parameter, corresponding to the asymptotic separation between the D8-branes, but we focus exclusively on the case where the two stacks are on opposite sides of circle and extend down to the tip of the cigar. }

\begin{figure}
\centering
\includegraphics{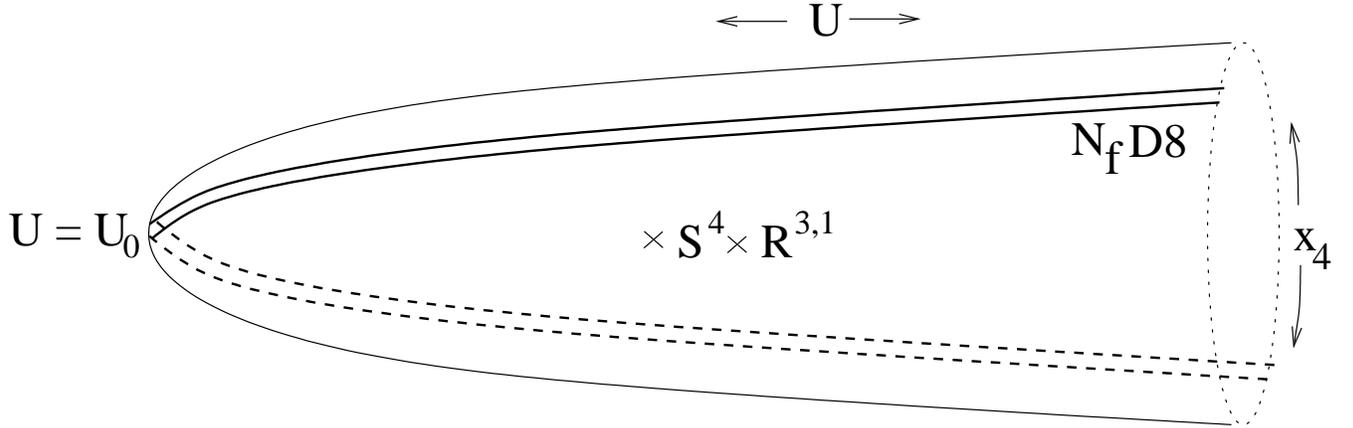}
 \caption{Type IIA string theory configuration for the Sakai-Sugimoto model.}
\end{figure}

For small values of $\lambda$, the scale $\Lambda_{QCD}$ where the running coupling becomes large is well below the field theory Kaluza-Klein scale, and the low-energy physics should be precisely that of pure $SU(N)$ Yang-Mills theory coupled to $N_f$ massless (fermionic) quarks.\footnote{For recent work on adding quark masses, see \cite{koji,evans}.} Unfortunately, in this limit, the dual gravity background is highly curved so we are not in a position to study it. For large $\lambda$ on the other hand, the gravity background is weakly curved, and so via classical calculations on the gravity side of the correspondence, it should be possible to map out the phase diagram of the field theory as a function of temperature and chemical potential and quantitatively determine properties of the various phases.

We do not expect our results to agree quantitatively with real QCD (both because the Kaluza-Klein scale is not well separated from $\Lambda_{QCD}$ for large $\lambda$ and because the classical calculations give only the leading terms in the $1/N$ expansion), but it would certainly be interesting to have a precise understanding of the phase diagram for a theory that is so similar to QCD. Indeed, at least some features of the phase structure and the qualitative behavior of certain transitions are likely to be the same as in QCD, and we might even hope for rough quantitative agreement for quantities that are relatively insensitive to $\lambda$ and $N_c$ (we will discuss one such quantity below) .
\vskip 0.1 in
\noindent
{\bf The transition to nuclear matter}
\vskip 0.1 in
Our focus in this paper will be on the part of the phase diagram for zero temperature and intermediate values of the baryon chemical potential. In real QCD, as we increase the chemical potential from zero, the equilibrium state (i.e. the ground state) continues to be the vacuum until some critical value of the chemical potential at which point it becomes advantageous for baryons to condense. A first approximation to this critical value is the baryon mass, since it is at this point where it becomes energetically favorable to add single baryons to the vacuum. In fact, the critical value is somewhat lower, since the baryons have a negative binding energy. At the critical value, we have a first order transition from the vacuum state to homogeneous nuclear matter with some minimal baryon density.\footnote{It is important to note that we are talking only about QCD and ignoring electromagnetism here. With electromagnetic interactions, the binding energy per nucleon is actually greater in iron nuclei than in homogeneous nuclear matter, so the transition to nuclear matter is preceded by a transition to solid iron.} The best estimate for the critical chemical potential comes by studying the masses of atomic nuclei as a function of nucleon numbers \cite{ntrans}. These are fit very well by the Weizsacker-Bethe semiempirical mass formula, which includes a term proportional to the number of nucleons,
\[
m_{vol} = - b_{vol} A
\]
to take into account the energy $-b_{vol}$ due to strong interactions of each nucleon in the interior of a nucleus with its neighbors plus the average kinetic energy per nucleon (non-zero due to Fermi-Dirac statistics). The best fit for this energy is
\be
\label{bvol}
b_{vol} = 16 \; {\rm MeV} \; .
\ee
Ignoring electromagnetic interactions, this gives the binding energy per nucleon in the limit of large nuclei, and thus should be a good approximation to the value for homogeneous nuclear matter just beyond the transition. Thus, the critical chemical potential for the transition to nuclear matter in QCD should be approximately
\[
\mu_c = M_B(1 - {b_{vol} \over M_B}) \approx M_B(1 - 0.017) \; .
\]
As we increase the chemical potential further, the baryon density and the energy per baryon will increase from their values just above the transition. Eventually we hit at least one more transition, to a phase characterized by quark-quark condensates \cite{rw}.

In this paper, we will study the physics of the transition to nuclear matter in the Sakai-Sugimoto model at large $\lambda$. Via classical calculations in the dual gravitational theory, we will be able to determine the critical chemical potential and calculate the baryon density $n_B(\mu)$ and the energy per baryon $e_B(\mu)$ for $\mu$ above the transition.
\vskip 0.1 in
\noindent
{\bf Expectations at large $N$}
\vskip 0.1 in
Since our gravity calculations will give results corresponding to the large $N_c$ limit of the field theory (with a fixed $N_f$), we should briefly recall the expectations for how baryons behave for large $N_c$ \cite{wittenlargeN}. In this limit, baryon masses and baryon-baryon interaction energies go as $N_c$, but the baryon size approaches a constant. Thus, we expect that both the baryon density above the transition and the binding energy per nucleon divided by the baryon mass to have a finite limit for large $N_c$. These properties indeed follow from our calculations.

One significant difference between the large $N_c$ theory and ordinary QCD is the expected behavior at asymptotically large values of the chemical potential. In both cases, we have attractive interactions between excitations on the Fermi surface that result in an instability, but the nature of the resulting condensates is different. Whereas for $N_c=3$ the instability is a BCS-type instability, believed to lead to a color superconductor phase, the dominant instability at large $N_c$ is toward the formation of ``chiral density waves'' \cite{dgr}, inhomogeneous perturbations in the chiral condensate with wave number of order twice the chemical potential. This suggests that the ground state for large $N_c$ QCD at large enough chemical potential is inhomogeneous, however the nature of the true ground state remains mysterious (see \cite{ss} for a recent discussion). We believe that our analysis sheds some light on this question, as we will discuss shortly.
\vskip 0.1 in
\noindent
{\bf Results for the Sakai-Sugimoto model}
\vskip 0.1 in
In the Sakai-Sugimoto model, a chemical potential for baryon number corresponds to a nonzero asymptotic value of the electrostatic potential on the D8-branes, equal on both asymptotic regions of the D8-brane. Generally, this potential behaves asymptotically (for radial coordinate $U$ to be described below) as
\[
A_0 \sim \mu_B + E {c \over U^{3/2}} + \dots \; .
\]
The baryon density $n_B$ is proportional to the asymptotic abelian electric flux $E$, so configurations with non-zero baryon density in the field theory correspond to D8-brane configurations with sources for the electric flux. These sources can be either string endpoints on the D8-branes which originate from D4-branes wrapped on the internal $S^4$ of the geometry \cite{wittenbaryon} or (for $N_f > 1$) configurations of the Yang-Mills field carrying instanton charge \cite{ssbaryon,yibaryon}. The latter can be thought of as the wrapped D4-branes dissolved into the D8-branes and expanding into smooth instanton configurations.
\newpage
\noindent
{\bf One flavor}
\vskip 0.1 in
For any value of chemical potential, we always have a trivial solution for which the electrostatic potential is constant on the D8-branes and the baryon density is zero. However we can also consider translation invariant configurations with a uniform baryon density. In the single flavor case, which we consider first, the bulk description of baryons is in terms of pointlike instantons, since there are no large instanton configurations in the abelian gauge theory of a single D8-brane. In this case, configurations with a uniform baryon density correspond to having some density of these pointlike instantons on the D8-brane. For a given value of the chemical potential greater than the critical value, we find some preferred distribution of charges on the D8-brane. The total baryon density for a given value of $\mu$ may be read off from the asymptotic value of the electric flux, and the result increases smoothly from $0$ above the critical chemical potential, approaching an asymptotic behavior $n_B \propto \mu^{5 \over 2}$. The charge distribution in the radial direction for a given value of $\mu$ represents the distribution of energies in the condensate of baryons in the field theory. In particular, the distribution has a sharp edge at some value of the radial coordinate which increases for increasing chemical potential, and this gives a bulk manifestation of the (quark) Fermi surface in the field theory.

For the single flavor case, the transition to nuclear matter is continuous, unlike QCD, but it may be expected that the single flavor case is different due to the absence of pions which usually play a crucial role in interactions between nucleons.
{\
\vskip 0.1 in
\noindent
{\bf Two flavors}
\vskip 0.1 in
In the case with $N_f > 1$, we can have nonsingular instantons on the $N_f$ coincident D8-branes, and the minimum energy configurations for large enough $\mu$ are should involve smooth configurations of the nonabelian gauge field carrying an instanton density. While we might expect this to be homogeneous in the field theory directions, we argue that there are no allowed configurations of the D8-brane gauge field that are spatially homogeneous in the three field theory directions such that the net energy density and baryon density in the field theory are both finite. Thus, any phase with finite baryon density is necessarily spatially inhomogeneous. This has a simple interpretation: it suggests that at large $N_c$, the nucleons retain their individual identities for any value of the chemical potential. Assuming that this holds true also for small $\lambda$ where the theory becomes 2 flavor QCD, this suggests that the chiral density wave instability of the quark Fermi surface in large $N_c$ QCD simply indicates that the quarks want to bind into nucleons even at asymptotically large densities. This is discussed further in section 5.

To avoid the complication of directly studying inhomogeneous configuration, we approximate these by certain singular homogeneous configurations, arguing that our approximation should become exact in the limit of large densities. Within the context of this approximation, we study the behavior of the system as a function of chemical potential.

Our model displays a first order transition to nuclear matter at some critical chemical potential that depends on the parameter $\lambda$, with the baryon density behaving as $n_B \propto \mu^3$ for large $\mu$. In the limit of large $\lambda$, the critical value approaches the baryon mass, so the binding energy per nucleon is a vanishing fraction of the baryon mass at large $\lambda$.\footnote{While this statement is derived in the context of our approximation, we argue that it should be true in the full model.}

For large but finite $\lambda$, we find the behavior
\[
\mu_c = M_B^0 (1 + {c \over \lambda} + {\cal O} \left({1 \over \lambda^{3 \over 2}} \right))
\]
where $M_B^0$ is the large $\lambda$ result for the baryon mass
\[
M_B^0 = {1 \over 27 \pi} M_{KK} \lambda N_c \; .
\]
On the other hand, the baryon mass for large but finite $\lambda$ is \cite{ssbaryon,yibaryon}
\[
M_B = M_B^0 (1 + {c' \over \lambda} + {\cal O} \left({1 \over \lambda^{3 \over 2}} \right)) \; .
\]
It is interesting that the result for the binding energy per nucleon at the threshold for nuclear matter formation,
\[
E_{bind} = M_B - \mu_c  \approx {N_c \over 27 \pi} M_{KK} (c' - c) \; ,
\]
is actually insensitive to the value of $\lambda$ for large $\lambda$. Since we also know that this binding energy approaches some constant value in the limit of small $\lambda$ (the large $N_c$ QCD result with two massless flavors), then assuming a smooth behavior at intermediate values of $\lambda$, we can treat the large $\lambda$ result as a prediction for the order of magnitude of the QCD result.\footnote{Another example with similar insensitivity to $\lambda$ for both large and small $\lambda$ is the free energy of ${\cal N}=4$ SUSY Yang-Mills theory. Here, it is indeed the case that the large $\lambda$ result for the free energy gives a good prediction of the order of magnitude of the the small $\lambda$ result (or vice versa).} Noting that $M_{KK} \approx \Lambda_{QCD}$ for large $\lambda$, the value of the binding energy per nucleon extrapolated to $N_c=3$ becomes
\[
E_{bind} = {1 \over 9 \pi} \Lambda_{QCD} (c'-c) \approx 7 \; {\rm MeV} (c'-c)
\]
In order to reliably compute the the numerical coefficients $c$ and $c'$, we require knowledge of the nonabelian analogue of the Born-Infeld action, and (in the case of $c'$) probably corrections to this involving derivatives of field strengths. However, assuming $c'-c$ is of order one,\footnote{We must also assume that our approximation scheme at least gets the right power of $\lambda$ in the correction to $\mu_c$.} we do obtain the same order of magnitude as the $QCD$ result (\ref{bvol}). We are not aware of any other methods to reliably estimate this binding energy from first principles, so it is possible that a more complete calculation in the Sakai-Sugimoto model would represent the most reliable analytic prediction of this quantity.
\vskip 0.1 in
\noindent
{\bf Outline}
\vskip 0.1 in
The remainder of the paper is organized as follows. In section 2, we review the Sakai-Sugimoto construction and collect various results necessary for our investigation. In section 3, we review the description of baryons in the Sakai-Sugimoto model and outline the basic approach for studying the theory at finite chemical potential. In section 4, we consider the single flavor case, calculating the baryon density as a function of chemical potential above the transition to nuclear matter. In section 5, we discuss the two flavor case, introduce our approximation, and set up a variational problem that determines the minimal energy configuration with a fixed baryon density (within our approximation). We then study the variational problem numerically for various values of chemical potential and baryon density to determine the critical chemical potential above which the minimum energy configuration has non-zero baryon density.
\vskip 0.1 in
\noindent
{\bf Related Work}
\vskip 0.1 in
Our work complements and extends various previous studies of the phase diagram for the Sakai-Sugimoto model. The behavior at finite temperature was analyzed in \cite{Aharony:2006da}. The behavior of the Sakai-Sugimoto model at finite chemical potential has also been discussed (with a different focus from the present paper) in \cite{Kim:2006gp,Horigome:2006xu,Yamada:2007ys}. Discussions of the finite density behavior in other holographic models of QCD include \cite{Kobayashi:2006sb, Kim:2007xi, Kim:2007em, Domokos:2007kt,gutperle}

While this paper was in preparation, the paper \cite{Bergman:2007wp} appeared, which has some overlap with the present work, in particular section 4.1.

\section{The Sakai-Sugimoto model}

The basic setup for the Sakai-Sugimoto model \cite{ss1} begins with the low-energy decoupling limit of $N_c$ D4-branes wrapped on a circle of length $2 \pi R$ with anti-periodic boundary conditions for the fermions \cite{wittenthermal}. Apart from $N_c$, this theory has a single dimensionless parameter
\[
\lambda = {\lambda_{D4} \over 2 \pi R} \; ,
\]
the four-dimensional gauge coupling at the Kaluza-Klein scale. Because of the antiperiodic boundary conditions, the adjoint fermions receive masses of order $1/R$ while the scalars get masses of order $\lambda/R$ due to one-loop effects. The coupling runs as we go to lower energies, becoming strong at a scale
\[
\Lambda_{QCD} \sim {1 \over R} e^{- c \over \lambda}
\]
for some numerical constant $c$. As pointed out by Witten \cite{wittenthermal}, for small $\lambda$, the dynamical scale $\Lambda_{QCD}$ is far below the scale of the fermion and scalar masses and the Kaluza-Klein scale, so the dynamics should be exactly that of pure Yang-Mills theory.

The field theory here is dual to type IIA string theory on the near-horizon geometry of the branes. The Lorentzian metric, dilaton, and four-form field strength are given by
\beas
ds^2 &=& \left({U \over R_4} \right)^{3 \over 2}(\eta_{\mu \nu} dx^{\mu} dx^{\nu} + f(U) dx_4^2) + \left({R_4 \over U} \right)^{3 \over 2}({1 \over f(U)} dU^2 + U^2 d \Omega_4^2)  \cr
e^{\phi}  &=& g_s \left({U \over R_4} \right)^{3 \over 4} \cr
F_4 &=& {2 \pi N_c \over \omega_4} \epsilon_4
\eeas
where $\omega_4$ is the volume of a unit 4-sphere, $\epsilon_4$ is the volume form on $S^4$, and
\[
f(U) = 1 - \left({U_0 \over U} \right)^3 \; .
\]
The $x_4$ direction, corresponding to the Kaluza-Klein direction in the field theory, is taken to be periodic, with coordinate periodicity $2 \pi R$, however, it is important to note that this $x_4$ circle is contractible in the bulk since the $x_4$ and $U$ directions form a cigar-type geometry.

The parameters $R_4$ and $U_0$ appearing in the supergravity solution are related to the string theory parameters by
\[
R^3_4 = \pi g_s N_c l_s^3 \qquad \qquad U_0 = {4 \pi \over 9 R^2} g_s N_c l_s^3
\]
while the four-dimensional gauge coupling $\lambda$ is related to the string theory parameters as
\[
\lambda = 2 \pi {g_s N_c l_s \over R} \; .
\]
In terms of the field theory parameters, the dilaton and string-frame curvature at the tip of the cigar (the IR part of the geometry) are of order $\lambda^{3 \over 2}/N_c$ and $\sqrt{\lambda}$, so as usual, supergravity will be a reliable tool for studying the infrared physics when both $\lambda$ and $N_c$ are large (in this case, with $N_c >> \lambda^{3 \over 2}$).

Note that this is opposite to the regime of $\lambda$ where we expect pure Yang-Mills theory at low energies. However, we may still learn about pure Yang-Mills theory by studying this regime, since many qualitative features of the theory remain the same and we might expect further that certain quantitative features may be relatively insensitive to the value of $\lambda$ (as for example with the free energy in ${\cal N}=4$ SYM theory).

\subsection{Adding fundamental matter}

Now that we have defined the adjoint sector of the theory, we would like to add fundamental quarks. We keep the number of quark flavors fixed in the large $N_c$ limit, but this means that the number of degrees of freedom in the fundamental fields (including the gauge field) is smaller than the number of degrees of freedom in the adjoint sector by a factor $N_f/N_c$. Thus, for $N_f$ fixed in the large $N_c$ limit, the influence of the fundamental fields on the dynamics of the adjoint fields should be negligible.\footnote{It would be quite reasonable to argue that we should keep $N_f/N_c$ fixed for large $N_c$ to obtain a theory that is most qualitatively similar to QCD, since then the number of degrees of freedom in the adjoint and fundamental sectors of the theory remain of the same order of magnitude for large $N$. However, this limit is much more difficult to study using supergravity, since then the back-reaction of the matter branes, to be described presently, must be taken into account.} In other words, what is known as the ``quenched approximation'' in QCD literature is exact in this limit. This implies that adding the additional matter does not modify the geometry, and indeed the construction of Sakai and Sugimoto (following earlier constructions) involves adding branes to the geometry which are treated in the probe approximation.

The Sakai-Sugimoto construction is motivated by the observation that the light open string modes living at a 3+1 dimensional intersection of D4-branes and D8-branes give rise to chiral fermion fields on the intersection without accompanying bosons. Thus, to the original D4-branes, which we can take to lie in the 01234 directions with the $x_4$ direction periodic, Sakai and Sugimoto consider adding a stack of $N_f$ D8-branes and a stack of $N_f$ anti-D8 branes separated at fixed locations in the $x_4$ directions and extended along the remaining directions. This configuration is unstable before taking a near horizon limit\footnote{This instability is actually absent in the case we consider the stacks sit at opposite sides of the circle}, nevertheless, one can obtain a stable configuration of the probe branes in the bulk geometry by fixing the asymptotic positions of the D8 and D8-bar stacks in the $x_4$ direction. The $x_4$ positions of the branes are free to vary as a function of the radial direction $U$ in the bulk of the geometry, and charge conservation implies that the two stacks necessarily join up in the interior of the geometry. Thus,  (in the zero-temperature situation that we are considering) we really have just a single set of D8-branes, bent so that the orientation in the two asymptotic regions is opposite (see figure 1).

The specific embedding of the D8-branes in the bulk depends on the asymptotic separation of the stacks (and also any distribution of matter on the branes), but we will focus exclusively on the case where the two asymptotic parts of the D8-brane stack sit at opposite sides of the D8 circle, in which case each side simply extends to the tip of the cigar along a line of constant $x_4$ as shown in figure 1. The corresponding field theory has all flavors massless.

\subsection{D8-brane action}

To understand the physics of the probe D8-branes, we will need the action for the worldvolume D8-brane fields in the background above. We will begin by discussing the action for a single D8-brane before discussing the nonabelian generalization.

The Born-Infeld action for the worldvolume D8-brane fields (in the case of a single brane) is
\[
S = -\mu_8 \int d^9 \sigma e^{-\phi} \sqrt{-\det(g_{ab} + \tilde{F}_{ab})}
\]
where
\[
\tilde{F} \equiv 2 \pi \alpha' F
\]
We also have a Wess-Zumino term
\[
S =  \mu_8 \int e^{\tilde{F}} \wedge \sum C \; .
\]
Here, only the $C_3$ term contributes. Noting that $F^3$ is the derivative of the five-dimensional Chern-Simons form, $\omega_5$ and integrating by parts, we get
\[
S = - \mu_8 \int F_4 \wedge \omega_5 \; .
\]
After integrating over the sphere, this gives
\be
\label{cs}
S = {N_c \over 24 \pi^2} \int \omega_5(A)
\ee
where $d \omega_5 = F \wedge F \wedge F$. For a single D8-brane, $\omega_5 = A \wedge F \wedge F$.

To simplify the Born-Infeld action, we can choose to identify the worldvolume and spacetime coordinates in the sphere and the field theory directions, and parameterize the profile of the brane in the $U$ and $x_4$ directions by $U(\sigma)$ and $X(\sigma)$ respectively (we will soon focus on the solution where $X(\sigma)$ is constant).

We will be interested only in time-independent configurations homogeneous and isotropic in the spatial directions of the field theory (which we label by indices $i,j,k$). The most general configurations we will consider will have non-zero $F_{\sigma i}$, $F_{ij}$, and $F_{0 \sigma}$, all functions only of $\sigma$.

Integrating the determinant from the sphere directions over the sphere, we get a factor
\[
{8 \over 3} \pi^2 R_4^3 U
\]
while the remaining five-dimensional determinant is
\[
-\det(g_{\mu \nu}+ \tilde{F}_{\mu \nu}) = -(G_{00} g_{\sigma \sigma} + \tilde{F}_{0 \sigma}^2 + g_{00} \tilde{F}_{\sigma i} (g + \tilde{F})^{ij} \tilde{F}_{\sigma j}) \det(G_{ij} + \tilde{F}_{ij})
\]
with
\[
g_{\sigma \sigma} = G_{44} \partial_\sigma X \partial_\sigma X + G_{uu} \partial_\sigma U \partial_\sigma U \; .
\]
Note that we are using $G_{IJ}$ here to refer to the spacetime metric and $g_{ab}$ for the worldvolume metric. The final result (in the Abelian case) is
\bea
\label{bi}
S_{DBI} &=& -{\mu_8 \over g_s} {8 \over 3} \pi^2 R_4^3 \int d^4 x d \sigma U \left\{ \left( \left({U \over R_4} \right)^{3 \over 2} g_{\sigma \sigma} - \tilde{F}_{0 \sigma}^2 \right)\left( \left({U \over R_4} \right)^3 + {1 \over 2} \tilde{F}_{ij}^2 \right) \right. \cr
&& \left. \qquad \qquad \qquad \qquad \qquad  \qquad \qquad \qquad +  \left({U \over R_4} \right)^3  \tilde{F}_{ \sigma i}^2 + ({1 \over 2} \epsilon_{ijk} \tilde{F}_{i \sigma} \tilde{F}_{jk})^2 \right\}^{1 \over 2}
\eea
This action is manifestly invariant under reparametrizations of $\sigma$. The nonabelian generalization of this action is known only up to $F^6$ terms. Up to order $F^4$, we symmetrize all of the nonabelian field strengths in expanding the square root and take an overall trace. However, this symmetrized trace prescription is known to fail beyond order $F^4$.

\subsection{Chemical potential for baryon charge}

We would like to study the theory at finite chemical potential for baryon charge or alternatively, the theory with a modified Hamiltonian density
\[
H = H + \mu B
\]
where $B$ is the baryon charge density operator
\[
B = B_L + B_R = \psi_L^\dagger \psi_L + \psi_R^\dagger \psi_R \; .
\]
This is equivalent to adding a term $ - \mu B$ to the action since there are no time derivatives in $B$. Turning on the operator $B$ in the boundary gauge theory with real coefficient $\mu$ should correspond to turning on some (real) non-normalizible mode in the gravity picture. From the original brane setup, we know that the operators $B_L$ and $B_R$ couple to the time-components of the $D8$ and $\bar{\rm D8}$ brane gauge fields respectively. We will see below that the equations of motion for these fields require them to approach some constant values in the UV part of the geometry. If we describe the probe branes as above with a single gauge field for the whole configuration, then we have two such constant values,
\[
A_\infty = A_0(\sigma = \infty)
\]
and
\[
A_{-\infty} = A_0(\sigma = -\infty)
\]
These two values give the chemical potentials for the operators $B_L$ and $B_R$.\footnote{We give an argument in appendix A to establish that $B_L$ and $B_R$ are turned on with the same sign if $A_\infty$ and $A_{-\infty}$ have the same sign.} Thus, to work at finite chemical potential for baryon number, we require that the value of $A_0$ in both asymptotic regions of the D8-brane approaches the constant $\mu_B$.

\subsection{Asymptotic solutions}

In the simple case where the D8-brane is at constant $x_4$ and we assume that only the electrostatic potential is turned on, the Born-Infeld action above reduces to
\be
\label{bitwo}
S_{DBI} = -{\mu_8 \over g_s} {8 \over 3} \pi^2 R_4^{3 \over 2} \int d \sigma d^4 x U^{5 \over 2} \left[{1 \over f(U)} \partial_\sigma U \partial_\sigma U - \partial_\sigma \tilde{A} \partial_\sigma \tilde{A} \right]^{1 \over 2}
\ee
The reparametrization invariance allows us to chose $U(\sigma)$ to be whatever we like. For a given choice of $U$, the equation of motion for $A$ away from any sources (which we assume are localized in the infrared part of the geometry) is
\be
\partial_\sigma \left( {\mu_8 \over g_s} {8 \over 3} \pi^2 R_4^{3 \over 2}  U^{5 \over 2} \left[{1 \over f(U)} \partial_\sigma U \partial_\sigma U - \partial_\sigma \tilde{A} \partial_\sigma \tilde{A} \right]^{-{1 \over 2}} \partial_\sigma \tilde{A} \right) = 0
\label{cons}
\ee
The quantity in round brackets is analogous to the conserved electric flux. Integrating and rearranging, and choosing $\sigma = U$ (valid for either half of the brane), we get
\be
\label{defE}
\partial_u \tilde{A} = {E \over \sqrt{f(U)(U^5 + E^2)}} \; ,
\ee
where $E$ is an integration constant proportional to the conserved flux. Solving this, we find
\beas
\tilde{A} &=& \tilde{A}_\infty - \int_U^\infty du {E \over \sqrt{F(u)(u^5 + E^2)}} \cr
&=& \tilde{A}_\infty + {2 \over 3} {E \over U^{3 \over 2}} + \dots
\label{asympt}
\eeas
valid in the region outside the sources. The constant $E$ is the normalizible mode of $A_0$ in the asymptotic solution, so the values of $E$ for the two sides of the brane correspond to the expectation values for $B_L$ and $B_R$ in the field theory.

In general, the sum of the $E$s for the two halves of the brane (times ${\mu_8 \over g_s} {8 \over 3} \pi^2 R_4^{3 \over 2}(2 \pi \alpha')$) is equal to the total charge density on the brane,
\[
{\mu_8 \over g_s} {8 \over 3} \pi^2 R_4^{3 \over 2} (2 \pi \alpha') (E_2 + E_1) = q
\]
If we fix $A_\infty=A_{-\infty}$ as we have argued corresponds to a chemical potential for baryon number, and we assume that the sources are symmetric under a reflection in the $\sigma$ direction, then for continuous $A_0$ we must have $E_1=E_2$, and
\be
\label{chargevsE}
{\mu_8 \over g_s} {8 \over 3} \pi^2 R_4^{3 \over 2} (2 \pi \alpha') E = q/2
\ee
Since the charge density in the bulk (divided by $N_c$) corresponds to the baryon density in the field theory, we obtain
\be
\label{densdef}
n_B = {\mu_8 \over g_s N_c} {16 \over 3} \pi^2 R_4^{3 \over 2} (2 \pi \alpha') E
\ee

\section{Baryons}

We have seen that configurations with non-zero baryon charge density (as measured by the asymptotic electric flux $E$) require sources for $A_0$ on the D8-branes. The basic source for $A_0$ is the endpoint of a fundamental string. In order to have some net charge, we need the number of string endpoints of one orientation to be unequal to the number of string endpoints of the other orientation. So we need a source for fundamental strings in the bulk. In our background, such a source is provided by D4-branes wrapped on $S^4$ \cite{wittenbaryon}. These necessarily have $N_c$ string endpoints, since the background D4-brane flux gives rise to $N_c$ units of charge on the spherical D4-branes, so we need $N_c$ units of the opposite charge (coming from the string endpoints) to satisfy the Gauss law constraint. Thus, we can get a density of charge on the D8-brane by having a density of D4-branes wrapped on $S^4$ in the bulk, with $N_c$ strings stretching between each D4-brane and the D8-brane.

In the case where we have $N_f>1$ D8-branes, there is another possible picture of the configurations with baryons \cite{ssbaryon,yibaryon}. To see this, note that a D4-brane / D8-brane system with four common worldvolume directions is T-dual to a D0-D4 system. In that case, it is well known that the D0-branes can ``dissolve'' in the D4-branes, where they show up as instanton configurations of the spatial non-abelian gauge field. Similarly, our baryon branes can dissolve in the D8-branes (if we have $N_f>1$) and show up as instantons. Indeed, the Chern-Simons term (\ref{cs}) gives rise to a coupling
\be
\label{cs2}
S = {N_c \over 8 \pi^2} \int A_0 \tr(F \wedge F)
\ee
between the instanton charge density and the abelian part of the gauge field, showing that instantons act as a source for the electrostatic potential on the branes.

The question of which of these two pictures is more appropriate is a dynamical one, but it turns out that the dissolved instantons give rise to a lower energy configuration since the electrostatic forces prefer the instanton density to be delocalized \cite{ssbaryon,yibaryon}.

\subsection{Baryon mass}

The baryon mass was estimated originally by Sakai and Sugimoto \cite{ss1} as the energy of a D4-brane wrapped on $S^4$ and located at the tip of the cigar. Since we will also need to know the potential energy for such branes, we briefly recall the calculation. Starting with the Born-Infeld action for a D4-brane wrapping $S^4$,
\[
S = -\mu_4 \int d^5 \xi e^{- \phi} \sqrt{-\det(g_{ab})}
\]
and integrating over the sphere, we get
\be
\label{dfourpotl}
S_{D4} = -{\mu_4 \over g_s} {8 \over 3} \pi^2 R_4^3 \int dt U(t)
\ee
as the velocity independent term in the action (the negative of the potential energy). The minimum energy occurs for $U=U_0$, and this gives the baryon mass
\[
M_B^0 = {\mu_4 \over g_s} {8 \over 3} \pi^2 R_4^3 U_0 = {1 \over 27 \pi} {1 \over R} \lambda N_c \;
\]
This agrees with the Yang-Mills action for a pointlike instanton configuration on the D8-brane \cite{ssbaryon}. Both of these calculations ignore the energy from the electric flux sourced either by the string endpoints coming from the wrapped D4-brane or by the instanton density. To take this into account, the authors of \cite{ssbaryon} and \cite{yibaryon} considered more general smooth instanton configurations with varying scale factor, inserting these into the Yang-Mills approximation to the D8-brane action. They found that the optimal size for the instanton behaves as $\lambda^{-{1 \over 2}}$, and that the baryon mass is
\[
M_B = M_B^0(1 + {c' \over \lambda})
\]
This method ignores the effects of the non-trivial geometry on the Yang-Mills configuration and also does not include effects from the $\alpha'$ corrections to the D8-brane effective action, which should be important, since for large $\lambda$, the instanton is small so that derivatives of the Yang-Mills field strength are large. Thus, as the authors point out, the numerical coefficient $c'$ should probably not be trusted. On the other hand, an analysis of the effects of Born-Infeld corrections \cite{ssbaryon} indicates that at least the power of $\lambda$ in the correction to the mass and in the instanton size should be reliable.

\subsection{Critical Chemical Potential}

We have seen that turning on a chemical potential in the gauge theory corresponds to including boundary conditions $A_0 = \mu$ for the two asymptotic regions of the D8-brane. For any $\mu$, one solution consistent with these boundary conditions is to have constant $A_0$ everywhere on the brane.  This represents the vacuum configuration in the field theory. However, beyond a certain critical chemical potential, this solution is unstable to the condensation of baryons.

The critical value of the chemical potential should not be larger than the baryon mass. At this value, a zero-momentum baryon has effectively negative energy in the modified hamiltonian, so it is advantageous to add baryons to the vacuum. If there were no interactions between the baryons, the critical chemical potential would be exactly the baryon mass. Note that even in the absence of interactions, the baryon density above the transition is limited by the Fermi statistics for the baryons for odd $N$ or in any case by the Fermi statistics of the quarks. The condensate will have occupied all states whose Fermi energy is less than the chemical potential. In this case, the baryon density will rise smoothly from zero above the critical chemical potential and the transition will be second order.

With short range repulsive interactions, the story would be qualitatively similar, with a slower growth in the baryon density as the chemical potential is increased. In QCD, however, we have attractive interactions, and this lowers the critical chemical potential below the baryon mass. With the repulsive interactions, there is a specific nonzero value of the baryon density for which the energy per baryon is lowest, and when the chemical potential is increased to this value the baryon density jumps from zero to this density.

In the next sections, we will study this transition to nuclear matter in the Sakai-Sugimoto model for one flavor (section 4) and two flavors (section 5). In the first case, it appears that the transition is second order, unlike QCD, while in the multi-flavor case, we find some evidence for a more realistic first-order transition.

\section{One flavor physics}

In this section, we study the physics of the Sakai-Sugimoto model at finite chemical potential in the simpler case of a single quark flavor. Here, we have only a single D8-brane in the bulk, and we can use the abelian Born-Infeld action for our analysis. Since the abelian gauge theory does not support large instantons, the wrapped D4-branes cannot dissolve into the D8-branes, so the baryons are pointlike charges on the D8-brane that source the electrostatic potential. For chemical potential larger than the baryon mass, it is favorable for some of these baryons to condense, and we would now like to determine the baryon density as a function of chemical potential for $\mu$ above the critical value.

\subsection{Localized source approximation}

As a first approximation, we make the simplifying assumption that all the pointlike instantons sit at $U=U_0$. More realistically, the charge should spread out dynamically, via electrostatic repulsion; we will include this effect in section 4.2.

In our simple approximation, the relevant action is the Abelian Born-Infeld action (\ref{bitwo}), together with the action taking into account the baryon masses and their interaction with the electromagnetic field on the brane.
\beas
\label{bithree}
S &=& -{\mu_8 \over g_s} {8 \over 3} \pi^2 R_4^{3 \over 2} \int dU d^4 x U^{5 \over 2} \left[{1 \over f(U)} - \partial_\sigma \tilde{A} \partial_\sigma \tilde{A} \right]^{1 \over 2} \cr
&&+ {n_B N_c \over 2 \pi \alpha'} \tilde{A}(U_0) - n_B M_B^0
\eeas
where the terms in the last line are the potential terms taking into account the energy from the charges in the electrostatic potential and the masses of the pointlike instantons.

To obtain the energy, we perform a Legendre transform, but it is convenient first to rewrite the first term in the second line as
\[
{n_B N_c \over 2 \pi \alpha'} \tilde{A}(U_0) = {n_B N_c \over 2 \pi \alpha'} \tilde{A}_\infty - \int dU {n_B N_c \over 2 \pi \alpha'} \tilde{A}'(U)
\]
since we will be holding $A(\infty) = \mu$ fixed. Performing the Legendre transform (which amounts to taking the negative of the action, since we are only looking at static configurations), and rewriting everything in terms of the electric flux (\ref{defE}), we find
\beas
{\cal E}_{flux} &=& 2 \cdot {\mu_8 \over g_s} {8 \over 3} \pi^2  R_4^{3 \over 2} \int_{U_0}^\infty dU { U^{5 \over 2} \over \sqrt{f}} ( \sqrt{1 + {E^2 \over U^5 }} - 1) -(\mu - \mu_c) n_B\cr
&=&  {\mu_8 \over g_s} {16 \over 3} \pi^2   R_4^{3 \over 2} U_0^{7 \over 2} h (e)-(\mu - \mu_c) n_B
\eeas
where we have defined $e = E/U_0^{5 \over 2}$ and
\[
h(e) = \int_1^\infty dx (\sqrt{x^5 + e^2} - x^{5 \over 2}){1 \over \sqrt{1-1/x^3}} \; .
\]
In the first term, we have included a factor of 2 to take into account the energy f
rom both halves of the D8-brane.

For $\mu > \mu_c$, the combined energy from the string endpoints (or Chern-Simons action) and the D4-brane mass (or Born-Infeld energy of the instantons) is negative and should be proportional to $n_B$, while the energy from the flux is a positive function of $n_B$ which behaves as $n_B^2$ for small $n_B$ and $n_B^{7 \over 5}$ for large $n_B$. Thus, there will be some positive value of $n_B$ where the total energy is minimized.

Defining
\[
\tilde{\mu} = {6 \pi \alpha' \mu \over U_0} \; ,
\]
so that $\tilde{\mu}=1$ corresponds to $\mu = M_B$, and using the relation (\ref{chargevsE}) between $n_B$ and $E$, the total energy may be written as
\[
{\cal E} = {\mu_8 \over g_s} {16 \over 3} \pi^2 R_4^{3 \over 2} U_0^{7 \over 2}\left(h(e) - {1 \over 3}(\tilde{\mu} - 1) e \right) ; .
\]
From this, we find that the energy is minimized when
\[
{1 \over 3} (\tilde{\mu} - 1) = h'(e) \; .
\]
This can be inverted to determine the relationship between $n_B$ (proportional to $e$) and $\mu$ above the transition. For small $\mu- \mu_c$, we find
\[
e \sim {1 \over \pi} (\tilde{\mu} - 1)
\]
so
\[
n_B \propto \mu - \mu_c \qquad \qquad {\rm small \;} \mu - \mu_c \; .
\]
For large $\mu$ we have
\[
e \sim 0.021 \tilde{\mu}^{5 \over 2}
\]
so
\[
n_B \propto \mu^{5 \over 2} \qquad \qquad {\rm large \;} \mu - \mu_c
\]

\subsection{Dynamical charge distribution}

The analysis of the previous section assumed that all charges were localized at $U=U_0$. Presumably, the charges would prefer to spread out dynamically. To take this into account, we can define a charge distribution $\rho_B(U)$ which we would like to determine. For a given $\rho$, the action is given in terms of a Lagrangian density
\[
{\cal L} = - C U^{5 \over 2} \left({1 \over f(U)} - \partial_\sigma \tilde{A} \partial_\sigma \tilde{A} \right)^{1 \over 2} + {N_c \over 2 \pi \alpha'} \tilde{A} \rho_B - {N_c \over 6 \pi \alpha'} U \rho_B \; .
\]
where
\[
C =  {\mu_8 \over g_s} {8 \over 3} \pi^2 R_4^{3 \over 2} \; .
\]
Here, the second term is the action arising from the string endpoints, while the third term takes into account the potential energy from the baryon masses (recalling that the action for a wrapped D4-brane at location $U$ is proportional to $U$).

For a given $\rho_B$, the electric flux is determined by solving the equation of motion for $\tilde{A}$,
\be
(2 \pi \alpha') \partial_U \left( {\mu_8 \over g_s} {8 \over 3} \pi^2 R_4^{3 \over 2}  U^{5 \over 2} \left[{1 \over f(U)} - \partial_U \tilde{A} \partial_U \tilde{A} \right]^{-{1 \over 2}} \partial_U \tilde{A} \right) = \rho_B(U) N_c
\label{cons}
\ee
This gives
\[
\rho_B(U) = {C (2 \pi \alpha') \over N_c} \partial_U E
\]
where we have defined an electric flux
\[
E(U) = U^{5 \over 2} ({1 \over f(U)} - (\partial_U \tilde{A})^2)^{-{1 \over 2}}\partial_U \tilde{A} \; .
\]
We can now reexpress all terms in the action in terms of $E$ and Legendre transform (which again amounts to switching the sign) to find the energy. We obtain
\beas
{{\cal E} \over 2C} &=&  \int_{U_0}^\infty dU \left[ {1 \over \sqrt{f}}(\sqrt{U^5 + E^2}-U^{5 \over 2}) + {1 \over 3} U \partial_U E \right] - \tilde{A}_\infty E_\infty\cr
\eeas
where we have included an extra factor of 2 in the denominator on the left side since we are integrating over only half the brane on the right side. To maximize this, we can first minimize over all $E(U)$ such that $E(U_0)=0$, $E(U \to \infty)= E_\infty$, and $\partial_U E>0$ to determine ${\cal E}(E_\infty, \mu)$. Then we can minimize over $E_\infty$.

Varying the energy functional with respect to $E$, we find that the energy functional is locally stationary if and only if
\be
\label{optimal}
{E \over (U^5 + E^2)^{1 \over 2}} = {\sqrt{f(U)} \over 3}
\ee
This satisfies $E=0$ for $U=U_0$ as desired but approaches arbitrarily large values for large $U$. On the other hand, our constraints $\partial_U E > 0$ and $E(U \to \infty) \to E_\infty$ imply that $E$ can never exceed $E_\infty$. It is straightforward to check that the local contribution to the energy from a point $U$ is a function of $E$ that decreases from $E=0$ to the optimal value (\ref{optimal}) and then increases again, so when the value (\ref{optimal}) exceeds $E_\infty$, the best we can do to minimize the energy is to set $E=E_\infty$. We conclude that the minimum energy configuration for fixed $\mu$ and fixed $E_\infty$ is
\be
\label{bestE}
\ba{ll}  E = {U^{5 \over 2} \over \sqrt{{9 \over f} - 1}} & U < U_{max} \cr E = E_\infty & U \ge U_{max} \ea
\ee
Here $U_{max}$ represents the extent of the charge distribution in the radial direction, and is related to $E_\infty$ as
\be
\label{EinffromU}
{E_\infty \over (U_{max}^5 + E_\infty^2)^{1 \over 2}} = {\sqrt{f(U_{max})} \over 3}\ee
We can now write the energy as a function of $E_\infty$, or more conveniently, $U_\infty$ as follows. We define a function $g(x)$ by
\[
g(x) = {x^{5 \over 2} \over \sqrt{{9 \over \tilde{f}(x)} - 1}}
\]
where
\[
\tilde{f}(x) = 1 - {1 \over x^3} \; ,
\]
and define
\[
H(x,g) = {1 \over \sqrt{f(x)}}(\sqrt{x^5+g^2}-x^{5 \over 2}) \; .
\]
Then in terms of $u = U_{max}/U_0$ and , the energy is given by
\[
{\cal E} = 2 C U_0^{7 \over 2} \left\{ \int_1^u dx H(x,g(x)) + \int_u^\infty H(x,g(u)) - {1 \over 3} \int_1^u g(x) dx + {1 \over 3} u g(u) - {1 \over 3} \tilde{\mu} g(u) \right \}
\]
where as in the previous section, we define
\[
\tilde{\mu} = {(6 \pi \alpha') \mu \over U_0} \; .
\]
We can now minimize this as a function of $u$. The result is
\[
\tilde{\mu} =  u + 3 \int_u^\infty dx \partial_g H(x,g(u))
\]
To compare with the results of the previous section, we note that (using (\ref{EinffromU})) the dimensionless variable $e$ proportional to the baryon mass is related to $u$ by
\[
{e  \over (u^5 + e^2)^{1 \over 2}} = {\sqrt{\tilde{f}(u)} \over 3} \; .
\]
From these, we find that for small $u-1$,
\[
\tilde{\mu} - 1 = c_1 (u-1)^{1 \over 2} \qquad \qquad c_1 \approx 1.814 \qquad \qquad {\rm small} \; u-1
\]
or
\[
e \sim {1 \over \pi} (\tilde{\mu}-1)
\]
where we have used (\ref{EinffromU}). Thus, for small $e$ we obtain the same result as in the previous approximation, with
\[
n_B \propto (\mu - \mu_c)
\]
for small $\mu - \mu_c$, where the critical value of $\mu$ is as before. For large $\mu$, we find
\[
\tilde{\mu} \to c_2 u \qquad \qquad c_2 \approx 1.697 \qquad \qquad {\rm large} \; u
\]
or
\[
e \sim 0.0942 \tilde{\mu}^{5 \over 2} \; .
\]
Again, we find that
\[
n_B \propto \mu^{5 \over 2} \; .
\]
Thus, the qualitative behavior of $n_B(\mu)$ is the same as in the simplified model of the previous section, though the numerical coefficients come out different. We also found the behavior of the energy density:
\[
{\cal E} \propto (\mu - \mu_c )^ 4
\]
for $(\mu - \mu_c )$ small, and
\[
{\cal E} \propto \mu^{7/2}
\]
when $\mu $ is large.

It is interesting that (in this approximation) the charge distribution has a sharp edge at $U=U_{max}$ which progresses further and further towards the UV in the radial directions as the chemical potential is increased. In the field theory picture, the radial direction represents an energy scale, so the charge distribution we find in the bulk should be related to the spectrum of energies for the condensed baryons. The edge of the distribution is then a bulk manifestation of the Fermi surface.

Since our large $N_c$ calculation does not distinguish between even and odd values of $N_c$, it is insensitive to whether or not the baryons are fermions or bosons. Thus, the Fermi surface that we see should probably be thought of as the quark Fermi surface. It is interesting that the fermionic nature of the quarks in the field theory arises in the bulk from the classical electrostatic repulsion between the instantons.

\section{Two massless flavors}

For $N_f=2$, the authors of \cite{ssbaryon,yibaryon} argued that single instantons on the D-brane prefer to grow to some finite size on the baryon in order to balance the electrostatic forces which tend to make the instanton spread out with the gravitational forces which prefer the instanton to be localized as much as possible near the IR tip of the D8-branes. From these considerations, we also expect that the minimum energy configurations with nonzero baryon density will involve some smooth configuration of the nonabelian gauge field on the D8-brane locally carrying an instanton density $\tr(F \wedge F)$. In this section, we consider such configurations.
\newpage
\noindent
{\bf The absence of homogeneous configurations}
\vskip 0.1 in
We first consider static, spatially homogeneous configurations, such that $A_\mu$ is translation invariant in the 3+1 directions of the field theory and rotationally invariant (up to a gauge transformation) in the three spatial directions (which we denote by an index $i$). The general configuration of the spatial gauge field with these symmetries is
\be
\label{instansatz}
A_\sigma = 0 \qquad \qquad A_i = {1 \over 4 \pi \alpha'} \sigma_i h(\sigma)
\ee
for an arbitrary function $h(\sigma)$. These give\footnote{We use conventions where $\{\sigma_i, \sigma_j\} = 2 \delta_{ij} \identity$ and recall that $\tilde{F} \equiv (2 \pi \alpha') F$.}
\be
\tilde{F}_{ij} = -{1 \over 4 \pi \alpha'} \epsilon_{ijk} \sigma_k h^2(\sigma) \qquad \qquad \tilde{F}_{i\sigma} = - {1 \over 2} \sigma_i h'(\sigma) \; .
\label{ansatz}
\ee
From these, we find that
\[
\tilde{F}_{i\sigma} \tilde{F}_{i\sigma} = {3 \over 4} (h'(\sigma))^2 \identity_{2 \times 2} \qquad \qquad {1 \over 2} \tilde{F}_{ij} \tilde{F}_{ij} = {3 \over (4 \pi \alpha')^2} h^4(\sigma) \identity_{2 \times 2}  \; .
\]
We see that unless both $h$ and $h'$ vanish for $\sigma \to \pm \infty$, the Yang-Mills action density integrated over $\sigma$ will diverge, corresponding to an infinite energy density in the field theory. On the other hand, we find
\[
(\tilde{F} \wedge \tilde{F})_{123 \sigma} = {1 \over 8 \pi \alpha'} h^2(\sigma) h'(\sigma) = {1 \over 24 \pi \alpha'} \partial_\sigma (h^3(\sigma)) \identity_{2 \times 2} \; .
\]
In order that we have a configuration with finite baryon density in the field theory, we require that this instanton density, integrated over the sigma direction be non-zero\footnote{To see this, note that the abelian electrostatic potential $A_0$ couples to $\tr(F \wedge F)$, so that the change in the action upon a constant shift in $A_0$ (corresponding to a change in the baryon chemical potential) is $\int d \sigma \tr(F \wedge F)$.}. But this requires that $h(\infty) \ne h(-\infty)$, and we have already seen that such a configuration will result in an infinite energy density in the field theory.

The apparent conclusion for the dual field theory is that {\it there are no spatially homogeneous configurations with finite non-zero baryon density and finite energy density}. Now, there certainly are non-homogeneous configurations with finite average energy density and finite average baryon density: we can simply take a periodic array of individual instantons. For large enough chemical potential (greater than the energy density divided by the baryon density), such configurations are favored over the vacuum, so we will certainly have a phase transition to a phase with nonzero baryon density as the chemical potential is increased. However, our observation suggest that this phase cannot be spatially homogeneous.
\vskip 0.1 in
\noindent
{\bf Interpretation of the inhomogeneity and origin of the chiral density wave}
\vskip 0.1 in
The inhomogeneity of nuclear matter is not unexpected, and indeed is what we have for real nuclear matter at low densities (e.g. in the interior of large nuclei). It simply reflects the fact that the individual nucleons retain their identities (and therefore that the baryon density is clumped\footnote{Quantum mechanically this would be reflected in the behavior of density-density correlation functions.}). What is perhaps surprising is that the inhomogeneity seems to have a topological rather than a dynamical origin from the bulk point of view, following from basic properties of instantons. It follows that even at arbitrarily high densities, the nuclear matter will be inhomogeneous, though the scale of the inhomogeneities should become shorter and shorter as the instantons pack closer and closer together. This suggests an interpretation of the DGR ``chiral density wave'' instability of the quark Fermi surface \cite{dgr} at asymptotically large chemical potential: that even at arbitrarily high densities, quarks in large $N_c$ QCD bind into distinct nucleons, in contrast to the quark matter phase with homogeneous condensates that we expect at large $\mu$ for finite $N_c$. This may be related to the property that the density of a baryon diverges for large $N_c$ and thus the baryon is more and more sharply defined in this limit.
\vskip 0.1 in
\noindent
{\bf Our approximation}
\vskip 0.1 in
The absence of homogeneous configurations with finite baryon density complicates the analysis of the phase transition and the properties of the nuclear matter phase. We will not attempt to study the inhomogeneous configurations directly here. Rather, we will describe an approach that approximates the inhomogeneous configurations with singular homogeneous configurations.

Our approach is motivated by the observation that in the limit of infinite baryon density, the bulk configuration should become homogeneous. Such homogeneous configurations are singular at the core, corresponding to a divergence of the instanton charge density. For example, we can have a self-dual configuration of the form (\ref{instansatz}) if we choose
\be
\label{sd}
h(\sigma) = {1 \over \sigma} \; .
\ee
This should arise from the limit of a periodic array of instantons for which the separation is taken to zero while adjusting the scale factors to yield a non-trivial configuration in the limit. We expect that some similar configuration\footnote{not necessarily self-dual since we are working with the D-brane effective action in a nontrivial geometry} should arise in our case as the minimum energy configuration in the limit of infinite chemical potential.

As we move away from infinite density, the minimum energy configuration will only be approximately homogeneous. We expect, however, that the averaged field strengths and instanton density should be qualitatively similar to those for the configuration (\ref{sd}) but with finite values at $\sigma=0$. This behavior can be achieved in a configuration of the form (\ref{instansatz}) for which $h$ is an odd function like (\ref{sd}) but with some finite limit at $\sigma = 0$. Such configurations are singular at $\sigma=0$, but we will ignore any effects associated with the singularity at $\sigma=0$ since we are using our configurations to approximate non-singular inhomogeneous configurations that do not have any pathologies at $\sigma=0$.\footnote{This is similar in spirit to replacing a nonsingular charge distribution with a localized singular distribution with the same multipole moments.} In particular, we might expect that our approximation becomes exact in the limit of infinite baryon density where we can have homogeneous configurations. We will find evidence below that supports the validity of this claim. More generally, we find results that are in accord with various physical expectations, providing further evidence for usefulness of our approximation.

\subsection{Energy density for approximate configurations}

We would now like to analyze the behavior of the model as a function of chemical potential in the approximation where we consider only configurations of the form (\ref{instansatz}), taking $h$ to be a monotonically increasing function for $\sigma > 0$ that takes some finite (negative) value at $\sigma=0$ and vanishes for $\sigma \to \infty$. In practice, we work with the action for half the brane, assuming that $h$ is an odd function so that all the field strengths are symmetric about $\sigma=0$. As we mentioned above, such configurations are singular at $y=0$ but we ignore any effects of the singularity, motivated by the expectation that the nonsingular contributions may provide a good approximation to the averaged quantities for the non-singular inhomogeneous configuration that we should really be studying.

The configuration of the spatial $SU(2)$ Yang-Mills field carries instanton density, and therefore acts as a source for the abelian electrostatic potential on the D8-branes. In order to determine the potential $A(U)$ for a given $h(U)$, we need the equation of motion for $A$, which should come from the non-abelian generalization of the Born-Infeld action (\ref{bi}) and the Chern-Simons action (\ref{cs2}).

As we have noted, the nonabelian generalization of the Born-Infeld action (\ref{bi}) is known only up to $F^6$ terms. In the absence of the full result, we will work with a naive ordering prescription in which we simply insert our ansatz into the abelian expression (\ref{bi}) and (noting that each product of $F$s above gives an identity matrix) evaluate the trace. This will give us results that are precisely correct in the limit where the field strengths are small and only the Yang-Mills terms in the action are important, but we should not trust numerical coefficients whose calculation depends on the higher order terms in the Born-Infeld action.

Inserting the ansatz (\ref{ansatz}) into (\ref{bi}), we find (in the $\sigma=U$ coordinates):
\be
\label{nonborn}
S_{DBI} = -{\mu_8 \over g_s} {16 \over 3} \pi^2 R_4^3 \int d^4x dU U \sqrt{({1 \over f(U)} - (\partial_U \tilde{A})^2 + {3 \over 4}(h'(U))^2)((U/R_4)^3 + {3 \over 4} {h^4 (U) \over (2 \pi \alpha')^2}) }
\ee
while the Chern-Simons term (\ref{cs}) gives:
\bea
S &=& {N_c \over 24 \pi^2} \int \tr(A \wedge F \wedge F) \cr
&=& {N_c \over 128 \pi^6 (\alpha')^4} \int dU \tilde{A} \partial_U (h^3(U)) \; .
\label{noncs}
\eea
If we define
\[
G = {1 \over f(U)} + {3 \over 4} (h'(U))^2
\]
and
\[
F = U \sqrt{(U/R_4)^3 + { h^4(U) \over (4 \pi \alpha')^2}}
\]
then the action takes the form
\[
S = -C \int dU F \sqrt{G - (\partial_U \tilde{A})^2} + \hat{k} \int \tilde{A} \partial_U(h^3)
\]
where
\[
\hat{k} = {N_c  \over 128 \pi^6 (\alpha')^4}
\]
and
\[
C = {16 \over 3} \pi^2 {\mu_8 \over g_s} R_4^3
\]
The equations of motion for the electrostatic potential $A$ are
\[
C \partial_U E = \hat{k} \partial_U(h^3)
\]
where
\be
\label{EfromA}
E = {F \partial_U \tilde{A} \over \sqrt{G - (\partial_U \tilde{A})^2}} \; .
\ee
From this, we conclude that
\be
\label{kEone}
\hat{k} h^3 = C (E - E_\infty)
\ee
where we have determined the integration constant by demanding that $h$  vanish as $U \to \infty$, as is required for finite energy configurations. Since $E$ vanishes by symmetry at $U=U_0$ (assuming that there is no delta function charge distribution at $U=U_0$) we see that the asymptotic value of $E$ is related directly to the value of $h$ at $U=U_0$ by
\be
\label{E0fromh0}
\hat{k} h_0^3 = - C E_\infty \; .
\ee
We may therefore rewrite (\ref{kEone}) as
\[
E = {\hat{k} \over C} (h^3 - h_0^3)
\]
Using this result, the electrostatic potential may be determined in terms of $h$ by inverting (\ref{EfromA}).

We may now write an expression for the energy density of a configuration for a given value of $h(U)$.

Starting with the actions (\ref{nonborn}) and (\ref{noncs}), we can derive the 3+1 dimensional energy density via a Legendre transformation as we did earlier. We find
\[
{\cal E} = C \int dU \left[ {F \sqrt{G - (\partial_U \tilde{A})^2}}-F_{h=0}\sqrt{G_{h=0}} \right] + \hat{k} \int \partial_U \tilde{A} (h^3-h_0^3) - \hat{k} \tilde{A}_\infty h_0^3
\]
where we have subtracted off the energy density of the unexcited brane such that the vacuum state is normalized to zero energy.
We can now rewrite the energy in terms of $h$, assuming that the equation of motion for $A$ is obeyed. We have first
\[
{\cal E} = C \left\{ \int dU (\sqrt{G(F^2 + E^2)} - F_0 \sqrt{G_0} \right\} - C \tilde{A}_\infty E_\infty
\]
Now writing $E$ in terms of $h$ as above, changing variables to $x = U/U_0$, defining
\be
\label{ydef}
y = - {\sqrt{3} \over 2} {h \over U_0} \;,
\ee
\[
\lambda_0 = {2 g_s N_c l_s \over 3 \sqrt{3}R} \; ,
\]
and
\be
\label{tmudef}
\tilde{\mu} = \sqrt{3} R \mu = {\lambda_0 \over 3} {\mu \over M^{\lambda = \infty}_B} \; ,
\ee
we finally have
\be
\label{bienergy}
{\cal E} = {C U_0^{7 \over 2} \over R_4^{3 \over 2} } \left[ \int_1^\infty dx \left\{ \sqrt{{1 \over \tilde{f}(x)}+(y'(x))^2}
\sqrt{x^5 + \lambda_0^2(x^2  y^4 + (y^3-y_0^3)^2)} - {x^{5 \over 2} \over \sqrt{\tilde{f}(x)}} \right\}- \tilde{\mu} y_0^3 \right]
\ee
Using the definition (\ref{ydef}), and the relations (\ref{E0fromh0}) and (\ref{chargevsE}), we find that $y_0$ is related to the baryon density by
\beas
n_B &=& {\pi \over 12 \sqrt{3}} \left({4 \over 9 \pi} {g_s N_c l_s \over R} y_0 \right)^3 {1 \over R^3} \cr
&=& {2 \over 27 \pi^2} \lambda_0^3 y_0^3 {1 \over R^3}
\eeas
Thus, minimizing this expression for $\tilde{\mu}=0$ and fixed $y_0$ will give the minimum energy density for a fixed baryon density, which we denote by
\[
{\cal E}_{min}(y_0)
\]
The energy density per baryon is then proportional to ${\cal E}_{min}/y_0^3$, and as we have argued above, the minimum of this gives the critical chemical potential. In the next section, we will analyze the functional (\ref{bienergy}), to obtain results for the behavior of ${\cal E}_{min}(y_0)$ and for the critical chemical potential.

\subsection{Results}

In this section, we discuss the evaluation of the baryon density for a given chemical potential based on minimizing the energy functional (\ref{bienergy}). Demanding that the functional is stationary under local variations of $y$ gives a second order differential equation for $y$. For a given initial value $y_0$ we find that there is a particular value of the initial slope $y_0'$ for which the solution approaches 0 as $x \to \infty$. For larger or smaller $y_0'$ the solution approaches positive or negative infinity respectively for $x \to \infty$, giving a diverging energy functional, so the minimum energy configuration must correspond to the solution with boundary condition $y \to 0$ at $x \to \infty$.

\subsubsection{Small baryon density}

We first study ${\cal E}(y_0)$ in the regime where the baryon density is small. Since the full energy at finite $\mu$ takes the form
\[
{\cal E}(y_0,\mu) = {\cal E}_{min}(y_0) - \tilde{\mu} y_0^3 \; ,
\]
it is important to determine the behavior of ${\cal E}_{min}(y_0)$ for small $y_0$. As long as the potential for $\mu=0$ is quadratic (or linear) for small $y_0$, we must have a first order transition to some finite baryon density at a critical chemical potential rather than a continuous transition where the baryon density increases gradually from zero. The results we obtain at small $y_0$ are also very robust (within our approximation), since here all field strengths and derivatives are small, and the incompletely known $\alpha'$ corrections in the D8-brane effective action are not important.

The terms in (\ref{bienergy}) coming from the Yang-Mills action are simply the leading order kinetic and potential terms,
\[
{\cal E} = {C U_0^{7 \over 2} \over R_4^{3 \over 2}} \left\{ \int_1^\infty dx ( {1 \over 2} x^{5 \over 2} \sqrt{\tilde{f}(x)} (y'(x))^2 + {1 \over 2} \lambda_0^2 {1 \over x^{1 \over 2} \sqrt{f(x)}} y^4) - \tilde{\mu} y_0^3 \right\}
\]
It is convenient to change variables to obtain a canonical kinetic term. Thus, we define $u$ such that
\[
{du \over dx} = {1 \over x^{5 \over 2} \sqrt{\tilde{f(x)}}} =  {1 \over \sqrt{x^5 - x^2}}
\]
Choosing $u=0$ to correspond to $x=1$, we have
\[
x(u) = \sec^{2 \over 3}({3 \over 2} u) \; .
\]
Note that $x=\infty$ corresponds to $u = \pi/3$, so we now have a finite domain, which is convenient for our later numerical methods. Dropping the overall constant and working at $\mu=0$ for now, we have
\be
\label{ymapprox}
{\cal \tilde{E}} = \int_0^{\pi \over 3} du ({1 \over 2} (y')^2 + {1 \over 2} \lambda_0^2 x^2(u) y^4)
\ee
Extremizing, this gives rise to the differential equation
\be
\label{smallde}
y''(u) = 2 \lambda_0^2 x^2(u) y^3(u)
\ee
As we discussed above, for a given $y(0) > 0$, solutions to this equation with slope larger or smaller than some critical value will approach positive or negative infinity as $u \to \pi/ 3$ and give rise to an infinite energy. The minimal energy configuration corresponds to the critical value of the initial slope for which the solution approaches zero at $u=\pi/3$. For $y_0 << 1/ \lambda$, the solution is linear to a good approximation, since taking
\be
\label{smallysol}
y(u) = y_0 ( 1 - {3 \over \pi} u)
\ee
we find that the right hand side of (\ref{smallde}) is small enough that even the maximum value of $y''$ integrated over the interval would only change $y'$ slightly.

Thus, for $y_0 << 1/ \lambda$, the energy is given by inserting (\ref{smallysol}) into (\ref{ymapprox}), and we find
\[
{\cal \tilde{E}}_{eff}(y_0) \sim {3 \over 2 \pi} y_0^2 + {\cal O}(\lambda^2 y_0^4) \qquad \qquad {\rm small \;} y_0
\]
Thus, the full energy ${\cal E}(y_0, \mu)$ is always positive for small enough $y_0$, and the transition to nuclear matter must be first order in our model.

While we can no longer trust the Yang-Mills approximation for large $y_0$ (of order $1/\sqrt{\lambda}$ or larger), it still interesting to look at behavior of the Yang-Mills terms in the energy functional in this regime. Continuing to use only the terms (\ref{ymapprox}), a numerical study suggests the asymptotic behavior
\[
{\cal \tilde{E}}_{eff} \approx {1 \over 3} \lambda_0 y_0^3
\]
Note that this asymptotic growth in the energy density as a function of $y_0$ is not enough to stabilize the baryon density to finite values for $\tilde{\mu}$ larger than value
\[
\tilde{\mu} = {1 \over 3} \lambda_0 \; .
\]
Comparing with (\ref{tmudef}), we see that this value corresponds precisely to $\mu = M_B^0$. Thus, we conclude that the $\alpha'$ corrections in the Born-Infeld action are essential for stabilizing the baryon density to finite values for large $\mu$, and that without these, the baryon density would diverge beyond a critical chemical potential that exactly coincides with the large $\lambda$ result for the baryon mass. In fact, we will see that at large $\lambda$ the Born-Infeld corrections only modify this critical chemical potential by terms of order ${1 \over \lambda}$.

\subsubsection{The critical chemical potential}

Now that we have demonstrated that there must be a first order phase transition to nuclear matter in our model, we would like to determine the critical value of $\mu$ above which a non-zero baryon density is favored, and the baryon density as a function of $\mu$ above this.
Thus, we repeat our numerical study from the previous section, but this time with the full energy functional. In this case, the differential equation for $y$ (using the same coordinates) is
\[
y'' = -{5 \over 2 x^6} (y')^3 {dx \over du} + ((y')^2+x^5)\partial_y \ln H -y' {dx \over du} (1 + {(y')^2 \over x^5}) \partial_x \ln(H)
\]
where
\[
H(x,y) = \sqrt{1 + \lambda^2 ({y^4 \over x^3} +  {1 \over x^5} (y^3 - y_0^3)^2)} \; .
\]
As before, the energy is minimized for a critical solution to this equation that approaches 0 at $u = \pi/3$.

Our results indicate that the energy ${\cal E}_{min}(y_0)$ behaves as a quartic function of $y_0$ for large $y_0$, so the Born-Infeld terms stabilize the baryon density to finite values for any value of $\mu$. As we have discussed, the critical value of the chemical potential beyond which a nuclear matter phase is favored is given by the minimum value of the energy per baryon. Specifically, we have
\[
\tilde{\mu}_c = {\rm min}_{y_0} {{\cal \hat{E}} \over y_0^3} ;
\]
We have numerically evaluated this critical chemical potential for large values of $\lambda$ ranging from $\lambda=10$ to $\lambda=3000$. Our data for $\mu_{crit}$ at large $\lambda$ are fit very well with a function
\be
\label{fit}
\mu_{crit} = M_B^{0}( A  + c \lambda_0^{-1} + {\cal O}(\lambda_0^{-{3 \over 2}}))
\ee
where the best fit values are
\[
A \approx 1.000 \pm 0.001 \qquad c \approx 8 \pm 2
\]
Thus, to very good accuracy, the critical value of the chemical potential approaches the baryon mass for large $\lambda$. Though our analysis using singular homogeneous configurations is an approximation, it is implausible that the almost exact agreement between the critical chemical potential and the baryon mass that we find here for large $\lambda$ is a numerical coincidence. A more plausible explanation is that the ratio of the critical chemical potential to the baryon mass does approach 1 in the limit of large lambda, and that our approximation gets this leading result correct. This is in accord with the expectation that our approximation should become exact in the limit of large baryon density, since as we will see below, the baryon density just above the transition does approach infinity as $\lambda$ becomes large.

Thus, we believe that a robust conclusion of our analysis is that the binding energy per nucleon for large $\lambda$ is a vanishing fraction of the baryon mass.

\subsubsection{The binding energy per nucleon}

To determine the actual value of the binding energy, we need to compare the subleading term in (\ref{fit}) with the subleading term in the baryon mass.

Even if our approximation is also correct for this subleading term, evaluating the coefficient $c$ here depends crucially on the higher order terms in the Born-Infeld action. Since we have used the abelian D8-brane action together with an ad-hoc ordering prescription in lieu of the unknown full result for the effective action, we expect that the numerical value here is not reliable,  However, the result that the correction is of order $\lambda^{-1}$ (rather than e.g. $\lambda^{-{1 \over 2}}$) should be robust.

Similarly, a correct calculation of $c'$  in the result
\[
M_B = M_B^{0}( 1  + {c' \over \lambda_0} + \dots)
\]
for the baryon mass discussed in section 3.1 probably requires more complete knowledge of the non-abelian effective action. However, recalling that the leading order result for the baryon mass is proportional to $\lambda$, we see that the result for the binding energy per nucleon $(M_B - \mu_c)$ is actually relatively insensitive to $\lambda$ for large $\lambda$. Since we also know that this binding energy approaches some constant value in the limit of small $\lambda$ (the large $N_c$ QCD result with two massless flavors), then assuming a smooth behavior at intermediate values of $\lambda$, we can treat the large $\lambda$ result as a prediction for the order of magnitude of the QCD result.\footnote{Another example with similar insensitivity to $\lambda$ for both large and small $\lambda$ is the free energy of ${\cal N}=4$ SUSY Yang-Mills theory. Here, it is indeed the case that the large $\lambda$ result for the free energy gives a good prediction of the order of magnitude of the the small $lambda$ result (or vice versa).}

Noting that $M_{KK} \approx \Lambda_{QCD}$ for large $\lambda$, the value of the binding energy per nucleon extrapolated to $N_c=3$ becomes
\[
E_{bind} = {1 \over 9 \pi} \Lambda_{QCD} (c'-c) \approx 7 \; {\rm MeV} (c'-c)
\]
As we have noted in the introduction, this is indeed of the same order of magnitude as the physical QCD result of $16MeV$ assuming that $c'-c$ is of order one.

\subsection{Baryon density above the transition}

We can also calculate the baryon density just above the transition. Our results suggest that just above the transition, the preferred value of $y_0$ for large $\lambda$ behaves like
\[
y_0 \to K \lambda^{-{1 \over 2}}
\]
for $K \approx 0.31$. This suggests that
\[
n_B R^3 \propto \lambda^{3 \over 2}
\]
as $\lambda$ is increased. This is consistent with the finding of Sakai and Sugimoto that the baryon size goes like $\lambda^{-{1 \over 2}}$.

For large chemical potential, the result that the $\mu=0$ energy density approaches $y_0^4$ for large $y_0$ implies that the baryon density minimizing ${\cal E}(y_0 \propto n_B^{1 \over 3}) - \mu n_B$ for large $\mu$ is
\[
n_B \propto \mu^3
\]
Also, the energy density as a function of $\mu$ for large $\mu$ behaves as
\[
{\cal E} \propto \mu^4
\]
Note that the powers here are those appropriate for free fermions. We would like to understand this point better.

\section*{Acknowledgements}

We would like to thank Krishna Rajagopal, Steve Shenker, Greg van Anders and Eric Zhitnitsky for helpful comments and discussions. We are especially grateful to David Mateos for collaboration in the early stages of this project. This work has been supported in part by the Natural Sciences and Engineering Research Council of Canada, the Alfred P. Sloan Foundation, and the Canada Research Chairs programme.

\appendix

\section{Holographic Dictionary}

Consider a configuration (before decoupling) with a D4-brane in the
01234 directions (with $x_4$ noncompact) and a D8-brane in the 012356789 directions, but bent in a U shape so as to intersect the D4-brane at two places along $x_4$. Locally, one of these intersections is a D8 and the other is a $\bar{\rm D8}$. Now, we are interested in the coupling between the D8-brane gauge field and the operators
\[
B_L = \psi_L^\dagger \psi_L
\]
at the one intersection and
\[
B_R = \psi_R^\dagger \psi_R
\]
at the other intersection. We define $\psi_L^\dagger$ and
$\psi_R^\dagger$ such that they create particles with positive charge on
the D4-brane, or physically, such that a test charge on the D4-brane is
repelled from both of these particles. In this case, the baryon number
operator is
\[
B = 1/N_c( \psi_L^\dagger \psi_L + \psi_R^\dagger \psi_R)
\]
and we have a coupling $a_0 B$ in the effective action where $a_0$ is the
time component of the D4-brane gauge field.

We would now like to understand how the two operators $B_L$ and $B_R$
couple to the D8 brane gauge field. To do this, we note that if we perform
a rotation by $\pi$ in the 1-4 directions, centered at the point on the
D4-brane between the two D8-branes, we get back to precisely the same
configuration, since the D4-brane does not change orientation, while the
D8 and $\bar{\rm D8}$ branes will switch orientation but also switch position.

Now, suppose we have a configuration with one $\psi_L^\dagger$ particle at
the D8 intersection. This repels a test charge on the D4-brane, so after
the rotation it is still a particle that repels a test charge, but now it
is a particle at the D8-bar intersection. It must therefore be a
$\psi_R^\dagger$ particle. Thus, a $\psi_L^\dagger$ particle is mapped to
$\psi_R^\dagger$ particle. Now, suppose that we have a test charge on the
D8-brane that is repelled by the particle in the initial configuration. In
the rotated configuration, this test charge will still be repelled (by the
$\psi_R^\dagger$ particle). Also, the test charge in the new configuration
will have the same sign as in the old configuration, since we have simply
performed a rotation. This means that if we describe the entire U-shaped
D8-brane using a single patch, both $\psi_L^\dagger$ and $\psi_R^\dagger$
particle will source electric fields pointing away from the D4-brane (or
both towards the D4-brane, depending on our convention).

This implies further that if we use a single field $A_0$ over the entire
D8-brane configuration, then the coupling of $A_0$ to $B_L$ at the D8
intersection will have the same sign as the coupling of $A_0$ to $B_R$ at
the D8-bar intersection. For the Sakai-Sugimoto setup, this implies that
if we want to turn on a chemical potential for baryon number (i.e. turn on
the operator $B$), we want to choose $A_0$ to have the same sign at
$\sigma= \infty$ as at $\sigma = - \infty$ (if we use the same gauge field
over the whole brane configuration).

\end{document}